\documentclass[apj]{emulateapj}

\def\gtorder{\mathrel{\raise.3ex\hbox{$>$}\mkern-14mu
             \lower0.6ex\hbox{$\sim$}}}
\def\ltorder{\mathrel{\raise.3ex\hbox{$<$}\mkern-14mu
             \lower0.6ex\hbox{$\sim$}}}

\usepackage{colordvi}
\usepackage{amsmath}
\usepackage{hyperref}

\usepackage{academicons}

\slugcomment{Accepted to PSJ}

\shorttitle{Didymos Daily Spectral Evolution}

\shortauthors{Polishook et al.}

\begin{document}

\title{Near-IR Spectral Observations of the Didymos System - Daily Evolution before and after the DART Impact, indicates Dimorphos originated from Didymos
}
\author{
D. Polishook\altaffilmark{1},
F.E. DeMeo\altaffilmark{2},
B.J. Burt\altaffilmark{2,3},
C.A. Thomas\altaffilmark{4},
A.S. Rivkin\altaffilmark{5},
J.A. Sanchez\altaffilmark{6},
V. Reddy\altaffilmark{7},
}

\altaffiltext{1}{Faculty of Physics, Weizmann Institute of Science, Rehovot 0076100, Israel, 
david.polishook@weizmann.ac.il}
\altaffiltext{2}{Department of Earth, Atmospheric, and Planetary Sciences, Massachusetts Institute of Technology, 77 Massachusetts Avenue, Cambridge, MA 02139 USA}
\altaffiltext{3}{Lowell Observatory, 1400 West Mars Hill Road, Flagstaff, AZ 86001, USA}
\altaffiltext{4}{Northern Arizona University, Department of Astronomy and Planetary Science PO Box 6010, Flagstaff, AZ 86011 USA}
\altaffiltext{5}{The Johns Hopkins University Applied Physics Laboratory, Laurel, MD, USA}
\altaffiltext{6}{Planetary Science Institute, Tucson, AZ, USA}
\altaffiltext{7}{University of Arizona, Tucson, AZ, USA}

\begin{abstract}

Ejecta from Dimorphos following the DART mission impact, significantly increased the brightness of the Didymos-Dimorphos system, allowing us to examine sub-surface material. We report daily near-IR spectroscopic observations of the Didymos system using NASA’s IRTF, that follow the evolution of the spectral signature of the ejecta cloud over one week, from one day before the impact. Overall, the spectral features remained fixed (S-type classification) while the ejecta dissipated, confirming both Didymos and Dimorphos are constructed from the same silicate material. This novel result strongly supports binary asteroid formation models that include breaking up of a single body, due to rotational breakup of km-wide bodies.

At impact time +14 and +38 hours, the spectral slope decreased, but following nights presented increasing spectral slope that almost returned to the pre-impact slope. However, the parameters of the $1~\mu m$ band remained fixed, and no "fresh" / Q-type-like spectrum was measured. We interpret these as follow: 1. The ejecta cloud is the main contributor ($60-70\%$) to the overall light during the $\sim40$ hours after impact. 2. Coarser debris ($\geq 100~\mu m$) dominated the ejecta cloud, decreasing the spectral slope (after radiation pressure removed the fine grains at $\leq10$ hours after impact); 3. after approximately one week, the ejecta cloud dispersed enough to make the fine grains on Didymos surface the dominating part of the light, increasing the spectral slope to pre-impact level. 4. a negligible amount of non-weathered material was ejected from Dimorphos’ sub-surface, suggesting Dimorphos was accumulated from weathered material, ejected from Didymos surface.

\end{abstract}


\section{Introduction and Motivation} \label{sec:intro}

NASA’s Double Asteroid Redirection Test (DART) provides a unique opportunity to study the composition of an asteroid's interior. While the mission main goal was to study how an asteroid can be deflected from its trajectory by a kinetic impact \citep{Cheng2018, Rivkin2021, Cheng2023, Daly2023, Thomas2023}, additional knowledge can be learned by spectrally studying the evolution of the ejecta plume following the impact.

Pre-impact calculations preformed by Steve Chesley (private communication) and based on analysis tools documented in \cite{Fahnestock2022}, suggested the brightness of the ejected material would overcome Didymos' brightness for the few tens of hours following the impact. Indeed, a few minutes after the impact, the brightness of Didymos-Dimorphos increased dramatically, at two phases: 1. a fast and gaseous ejecta cloud containing Sodium and Potassium \citep{Shestakova2023}, and moving contra to the DART velocity vector in a speed of 1.5-1.7 km/sec, that lasted for a few minutes \citep[e.g.][]{Shestakova2023, Graykowski2023}; 2. a slow long lasting ejecta cloud around the body that formed a set of tails which changed in number, direction, and length during the months after the impact \citep[e.g.,][]{Li2023, Moreno2023}. The spectral observations described here were designed to measure the evolution of the slow ejecta cloud.

The significant increase in brightness due to the large cross section of the ejecta allows a unique observation of the material from which a satellite of an asteroid consists. Several models \citep{Scheeres2007, Walsh2008, Sanchez2011} suggest that asteroids with diameters smaller than a few tens of km can disrupt as their fast rotation applies tension on their weak internal strength, resulting in ejected material that goes into orbit and eventually accumulates into a satellite. There is evidence that such a process acted on Didymos: it has a fast spin of $P=2.2593\pm0.0001~h$ \citep{Pravec2006}; a spherical shape with an equatorial bulge \citep[][]{Naidu2020} that is a common feature among binary asteroids \citep[e.g.][]{Naidu2015}; and a relatively small diameter of $\sim780~m$ \citep{Scheirich2022} that is more sensitive to an accelerating spin due to thermal forces ({\it{i.e.}} the YORP effect) and stronger centrifugal forces to cause breaking \citep[e.g.,][]{Vokrouhlicky2015}.

A shared origin of Didymos and Dimorphos suggests a similar composition for the bodies. This is supported by the fact that asteroids belonging to the same collisional family usually present similar taxonomies \citep[e.g.,][]{Masiero2015}. Therefore, confirming an equivalent taxonomy for both Didymos and Dimorphos supports the rotational-disruption origin models for Dimorphos.

In addition, the fast velocity impact of DART, lifting material from layers tens or more cm thick, had the potential to expose sub-surface material \citep[e.g.,][]{Hirabayashi2022}. Such material theoretically has not been exposed to space weathering processes such as impacts of dust, solar particles and cosmic rays \citep[e.g.][]{Chapman2004, Clark2002}. Since the timescale of space weathering is short ($\sim10^6 ~y$ and shorter; e.g.,  \citealt{Vernazza2009, Nesvorny2010}), most asteroids’ surfaces are expected to be weathered, excluding those that recently experienced some infrequent events that excavate "fresh" material from internal layers (mechanisms such as collisions, planetary encounters, rotational-breakup etc. \citep[e.g.,][]{DeMeo2023}. The fact that non-weathered ({\it{a.k.a.}} "fresh") asteroid surfaces are observed (Q-types, \citealp{Binzel2010, Rivkin2011, Thomas2011, Thomas2012}) among the many weathered asteroids (S-types, both types represent L to LL ordinary chondrite meteorites; \citealp{DeMeo2022, Binzel2019}), supports the notion that fresh material can be exposed and even cover the entire asteroid surface.

Fresh material is recognized in the near-IR range by shallow spectral slope and a wider and deeper $1 \mu m$ absorption band (Q-type spectral features), compared to the higher spectral slope that S-type asteroids, such as Didymos, present \citep{deLeon2010}. Samples collected from the Moon \citep[e.g.][]{1970Sci...167..745H} and from the asteroid 25143 Itokawa \citep[e.g.][]{Abe2006, Hiroi2006} showed that the weathered layer is extremely shallow, in the range of a few nano- to micro-meter wide. Since material from deeper layers was excavated by the DART impact (on the scale of meters, as predicted by e.g., \citealp[]{Raducan2022, Stickle2022}) the ejecta originated from these layers had the potential to present non-weathered reflectance spectrum. Supporting this claim is the outburst event of the active asteroid P/2010 A2 \citep{Jewitt2010}, that was probably formed by a collision, similar in scale to the DART impact. Both \cite{Kim2012} and \cite{Jewitt2013A2} noticed color variation between the nucleus and the tail, attributing it to a probable difference between weathered surface to a fresh ejecta. Another active body, (6478) Gault, that produced a long tail during multiple times, was found to be an S-complex inner main belt asteroid, that temporarily (March 31, 2019) presented a reflectance spectrum with a low spectral slope  \citep{Marsset2019}. Classification differences between members of asteroid pairs were suggested as evidence for fresh material exposed by rotational disruption \citep{Polishook2014}. We should note that the level of weathering / reddening of the spectrum is a complex phenomenon, that is the function of many parameters such as the olivine-pyroxene ratio, the amount of exposure to weathering agents, and the surface structure (\citealt{Brunetto2015} and references within). Ordinary chondrites with less olivine, such as H-, and L-types will probably present less differences between weathered and non-weathered spectra. Therefore, in this case, it is harder to confirm or reject the exposure of non-weathered material.

Alternatively, these spectral features (low slope, wider and deeper $1~\mu m$  band) might be the results of other phenomena, such as composition and grain size. For example, silicate minerals with higher level of olivine compared to pyroxene have wider $1~\mu m$ band. Asteroids with such features were found among the Flora family of asteroids in the inner main belt \citep[e.g.][]{Vernazza2008, Binzel2019}, and the Koronis family of asteroids in the outer main belt \citep[e.g.][]{Rivkin2011, Thomas2011}, suggesting the Q-type asteroids are mostly originated from these specific groups. Shallower spectral slopes were suggested as the results of light reflected from surfaces with large grains ($>$100$\mu$m), as derived by laboratory irradiation experiments \citep{Hasegawa2019}. This result suggests that Q-type asteroids have standard weathered surfaces dominated by large grains. Therefore, careful measurements of spectral changes over time following the DART impact can shed light on the true source of any observed variation in the spectral shape and slope.

The small separation between Didymos and Dimorphos of $\sim 1~km$, had the potential to result with the ejecta of Dimorphos reach or even cover Didymos \citep[e.g.][]{Fahnestock2022, Hirabayashi2022, Moreno2022}. Since it is impossible to resolve the two bodies from the Earth, this scenario could be confirmed if the overall reflectance spectrum of the system would have change after the impact, and then remain fixed with time. In a case of only a temporary spectral change, the change is most probably due to the appearance of the impact ejecta and its following dispersion with time.

\section{Observations and Reduction}
To measure the spectral evolution of the extending ejecta cloud after the impact, we conducted daily observations of the Didymos system for a week from Maunakea using NASA's 3m IRTF telescope, on the same time of the day (see Table~\ref{tab:Observations}). Observations started on the night before the DART impact, in order to obtain a spectral baseline. We used IRTF/SpeX's \citep{Rayner2003} low resolution mode and acquired spectroscopic observations over $0.8$ to $2.45~\mu m$, a range suited for the compositional characterization of silicate-rich bodies. Using a long slit of $0.8''$-wide, that was synched with the parallactic angle, we observed in repeated A-B-B-A nod pattern in order to reduce the background. Exposure times of 120 seconds were used per image and each night we collected between 14 to 22 images. Guiding was conducted by keeping the bright target in the slit, while the telescope was tracking at Didymos' rate of motion in the sky. We used the MIT Optical Rapid Imaging System (MORIS; \citealp{Gulbis2011}), a camera assembled behind a dichroic window and collects the optical light, for guiding in broadband colors (V Johnson, SDSS gr). This observing and reduction protocol followed the one used by the MITHNEOS survey \citep[e.g.][]{Binzel2019, Marsset2022}, the most prolific survey by which most asteroids have been classified with, in order to reduce systematic errors.

Reduction of the raw SpeX images follow the procedures outlined in \cite{DeMeo2009} and \cite{Binzel2019}. This includes flat field correction, sky subtraction, manual aperture selection, background and trace determination, removal of outliers, and a wavelength calibration using arc images. An atmospheric transmission (ATRAN) model \citep{Lord1992}, was used to model and remove telluric lines. The spectrum was divided by a spectrum of a nearby star, HD25156 (RA = 03:58:39.277, Dec = -31:35:42.21, with angular distance from Didymos of 9 degrees on Sep 26, through a minimum of 53 arcmin on Sep 29, to 8 degrees on Oct 2), with solar colors\footnote{Respective solar colors: $B-V=0.653\pm0.005$, $V-I=0.702\pm0.01$ \citep{Ramirez2012}; V-J=1.198, V-H=1.484 \citep{Casagrande2011}} B-V=0.638, V-I=0.716, V-J=1.208, V-H=1.486 \citep{Wenger2000SIMBAD}, to derive the relative reflectance of the asteroid. The main challenge of the reduction was to deal with the high airmass of the Didymos system from Maunakea during the impact week ($1.5<AM<1.7$). To mitigate large systematic errors, we observed Didymos at the same hour at each of the program's nights (13:00-14:00 or 13:00-14:30 UT) in order to compare reflectance spectra taken at similar airmass and parallactic angle. We directed the slit towards the parallactic angle and calibrated the spectra against the same solar analog as near as possible to Didymos. For the analysis we use only the measurements taken between 0.8 to 2.45 $\mu m$, which is the standard and best tested wavelength range of SpeX \citep{Marsset2020}. All the images taken per night were co-added to present a single daily reflectance spectrum.

\begin{deluxetable*}{llccclllll}
\tablecolumns{10}
\tablewidth{0pt}
\tablecaption{\label{tab:Observations}
Observations Circumstances}
\tablehead{
\colhead{Night}    &
\colhead{Time}    &
\colhead{Images Num.} &
\colhead{AirMass} &
\colhead{Parallactic Angle} &
\colhead{R} &
\colhead{$\Delta$} &
\colhead{$\alpha$} &
\colhead{V-Mag.} &
\colhead{Solar analog} \\
\colhead{}       &
\colhead{UT}       &
\colhead{}       &
\colhead{}       &
\colhead{$^o$} &
\colhead{AU} &
\colhead{AU} &
\colhead{$^o$} &
\colhead{} &
\colhead{}
}
\startdata
Sep 26, 2022 & 13:23-14:30 & 22 & 1.68 & 0 & 1.047 & 0.076 & 53 & 14.55 & HD25156 \\
Sep 27, 2022 & 13:14-13:52 & 16 & 1.65 & 355 & 1.044 & 0.075 & 54 & 14.56 & HD25156 \\
Sep 28, 2022 & 13:16-13:54 & 14 & 1.62 & 353 & 1.042 & 0.074 & 56 & 14.57 & HD25156 \\
Sep 29, 2022 & 13:21-14:30 & 22 & 1.60 & 358 & 1.040 & 0.073 & 57 & 14.58 & HD25156 \\
Sep 30, 2022 & 13:16-13:53 & 16 & 1.58 & 349 & 1.037 & 0.072 & 59 & 14.60 & HD25156 \\
Oct 01, 2022 & 13:19-13:53 & 14 & 1.55 & 347 & 1.035 & 0.072 & 60 & 14.62 & HD25156 \\
Oct 02, 2022 & 13:09-13:52 & 18 & 1.53 & 347 & 1.033 & 0.071 & 62 & 14.65 & HD25156
\enddata
\tablecomments{Observations Circumstances: Date, time UT, number of images, air mass, parallactic angle, Solar-asteroid distance, Earth-asteroid distance, phase angle, visible magnitude, Solar analog. Geometric data taken from JPL's Horizons}
\end{deluxetable*}

\section{Results and Analysis}

\subsection{Spectral Slope Variation}
\label{sec:slopeVariation}

\begin{figure}
	\centering
             \includegraphics[width=8cm]{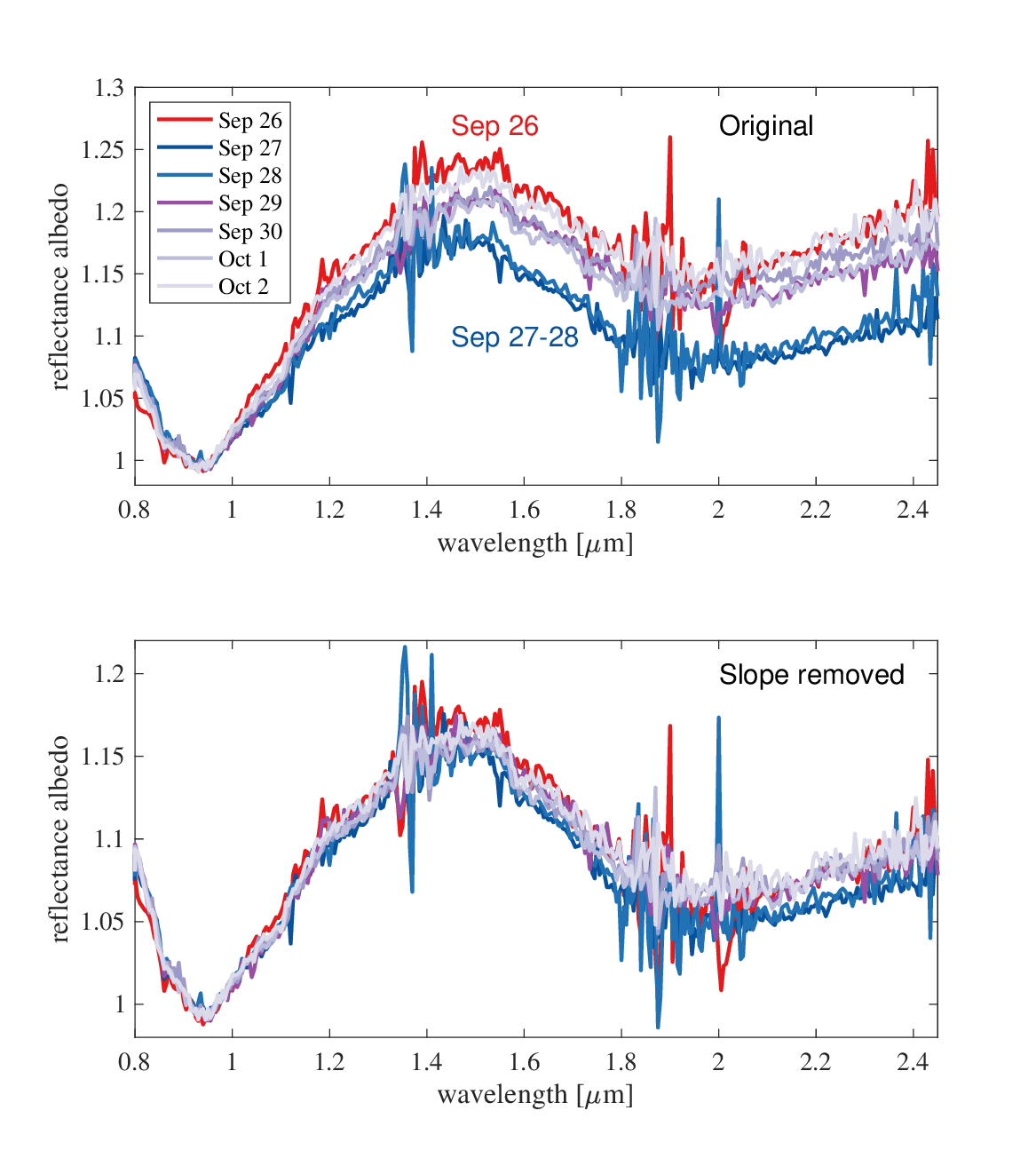}

	\caption{
	top panel: Seven near-IR reflectance spectra representing a daily measurement of the Didymos system. While all spectra have similar features, the immediate post-impact spectra (Sep 27, Sep 28) have spectrum with lower spectral slope (dark blue lines), compared to spectra from later dates (light purple lines), and compared to the pre-impact spectrum (Spec 26, red line. The colors representing the spectrum of each day remind fix through the paper). bottom panel: After removing the spectral slope using Eq.\ref{eq:DeWeathering}, there is a good match between all seven reflectance spectra, suggesting no change of material is present. The minimum value of a 6-order fit to the $1~\mu m$ absorption band was used to normalized all spectra for display. We conclude the slope variation is due to larger grains dominating the ejecta at the first $\sim40$ hours after the impact (see text for details).
	}

	\label{FIG:Spec_Sep26Oct2}
\end{figure}

The seven reflectance spectra, collected one per day, have SNR of roughly 100 through the entire range of wavelength, excluding the windows of telluric lines where the SNR deteriorates to 20. This allows us to conduct a detailed comparison between the spectra. Spectra are presented at the top panel of Fig.~\ref{FIG:Spec_Sep26Oct2}.

All the reflectance spectra are very similar in terms of features. After removing from each spectrum its slope using \cite{Brunetto2006} power-low equation (see text below and the bottom panel of Fig.~\ref{FIG:Spec_Sep26Oct2}), the median value of the standard deviation of the residuals after subtracting the post-impact spectra from the pre-impact spectrum (Sep 26 spectrum) is $1.75\%$.
 
The most striking difference is the spectral slope of the two reflectance spectra that followed the impact at $T_0+14$, $T_0+38$ hours, (where ${T_0}$ is the impact time Sep 26, 2022, 23:14:24 UT, \citealt{Daly2023}), compared to the spectral slope of the pre-impact and post-impact $T_0+62$ hours and afterwards. The slope value comparisons are summarized in Table~\ref{tab:Analysis} were we list two sets of slopes, one measured from 0.8 to 2.45 $\mu m$ and the other is measured from 0.8 to 1.5 $\mu m$. In both cases the decrease is on the order of tens of percentage. It should be stressed that spectral slopes can change due to observational circumstances, mainly due to the airmass of the target, the solar analog star and the distance between them \citep{Marsset2022}, and the phase angle \citep{Sanchez2012} (weather conditions and parallactic angle were not found to contribute to systematic error of the slope; \citealt{Marsset2022}), and therefore we tried to eliminate any such variation in the observation, by observing daily at a similar airmass, and normalizing with the same solar analog at a very close angular distance from the Didymos system. The phase angle changed very little over the duration of our observations (53 degrees at Sep 26, increasing at $\sim 1$ degree per day), and its slow rise is not correlated with the decrease of the spectral slope at Sep 27-28 nor its later increase. We have no reason to think that a systematic error should appear at the two days after the impact, then disappear on the following four nights. \cite{Marsset2022} analyzed 20 years of asteroid reflectance spectra and found a spectral slope uncertainty of $4.2\%~\mu m^{-1}$, which is significantly lower than the $\sim70\%$ difference between the spectral slope measured on Sep 26 (pre-impact), to that measured on Sep 27 (post-impact) - values of 0.082 and 0.026 $\mu m^{-1}$ respectively (slope here were measured on the entire wavelength range of 0.8 to 2.45 $\mu m$). However, uncorrelated systematic errors do occur and they can affect spectral slopes. Our analysis below is done under the assumption, based on the arguments given above, that this spectral slope represents a real physical behavior following the DART impact.

\begin{figure}
	\centering
             \includegraphics[width=8cm]{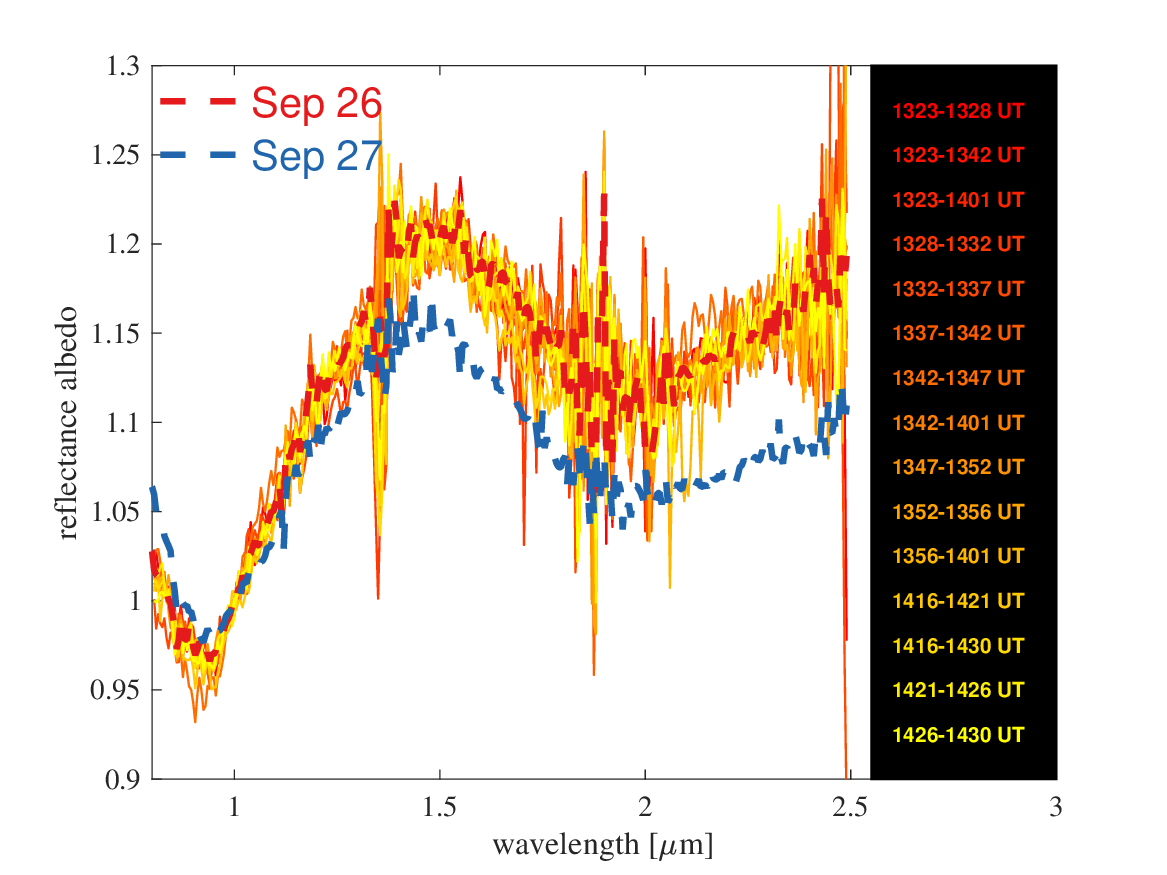}

	\caption{
	Reflectance spectra of each pair of images, or four images, taken on Sep 26, 2022, between 13:23 to 14:30 UT, that represent half of Didymos rotation (thin, colorful lines). The combined spectra of Sep 26, is also plotted (thick, red dash-line, the same line appear in Fig.~\ref{FIG:Spec_Sep26Oct2}). Post-impact reflectance spetrum (from Sep 27, in thick, blue dash-line) of the Didymos system is presented as a reference to a much smaller spectral slope. We found no variation within the scattering that is equivalent to the variation observed between spectra of Sep 27, Sep 28, to the rest of the spectra.
	}

	\label{FIG:Spec_Sep26Only}
\end{figure}

Variation of the reflectance spectra of the Didymos system might occur due to a color variation on Didymos surface, regardless of the DART impact. \cite{Ieva2022} reported on slope variation at the visible wavelenth range, observed before the impact (on February 2021). However, they could not correlate this variation with Didymos rotation phases. In order to test such a scenario, we compared the reflectance spectra of each couple of images, or four images, taken on Sep 26, between 13:23 to 14:30 UT, that represent half of Didymos synodic rotation ($P=2.2593\pm0.0001~h$, \citealp{Pravec2006}). We found no variation that is equivalent to the variation observed on the Sep 27-28 spectra compared to the pre-impact spectra and the Sep 29-Oct 2 spectra (Fig.~\ref{FIG:Spec_Sep26Only}).

In addition, we calculated the rotational phase of Didymos at mid-observation of each of the seven observing nights, relative to an arbitrary time, and check for any correlation between rotation phase and slope. This test failed to find any correlation since the two nights with shallower spectral slopes (Sep 27-28) were observed at opposite rotational phases (0.4 and 0.77) compared to one another, while the other spectra were taken at varied rotational phases (0.32, 0.18, 0.64, 0.28 0.86, at Sep 26, 29, 30, Oct 1, 2, respectively).

We also observed the Didymos system at other pre-impact dates during 4, 5 and 9 of August, but Didymos' faintness ($Vmag\sim17$) resulted in a scatter that is higher than the reported change in spectral slope (SNR less than 20 outside of the telluric windows).

Another possible spectral variation, not related to the DART impact, could occur in a theoretical case of large spectral variation on Dimorphos' surface. However, this is very unlikely to affect the overall spectra, since the surface ratio between Dimorphos and Didymos is $(0.17 km/0.78 km)^2=0.05$ (\citealt{Scheirich2022}) making any variation in Dimorphos spectrum insignificant compared to the light reflected from Didymos. Indeed, \cite{Ieva2022} observed the Didymos system during an eclipse event, and they did not detect any spectral variation between measurements within and outside the eclipse.

\subsection{Classification}
\label{sec:classification}

\begin{figure}
	\centering
             \includegraphics[width=8cm]{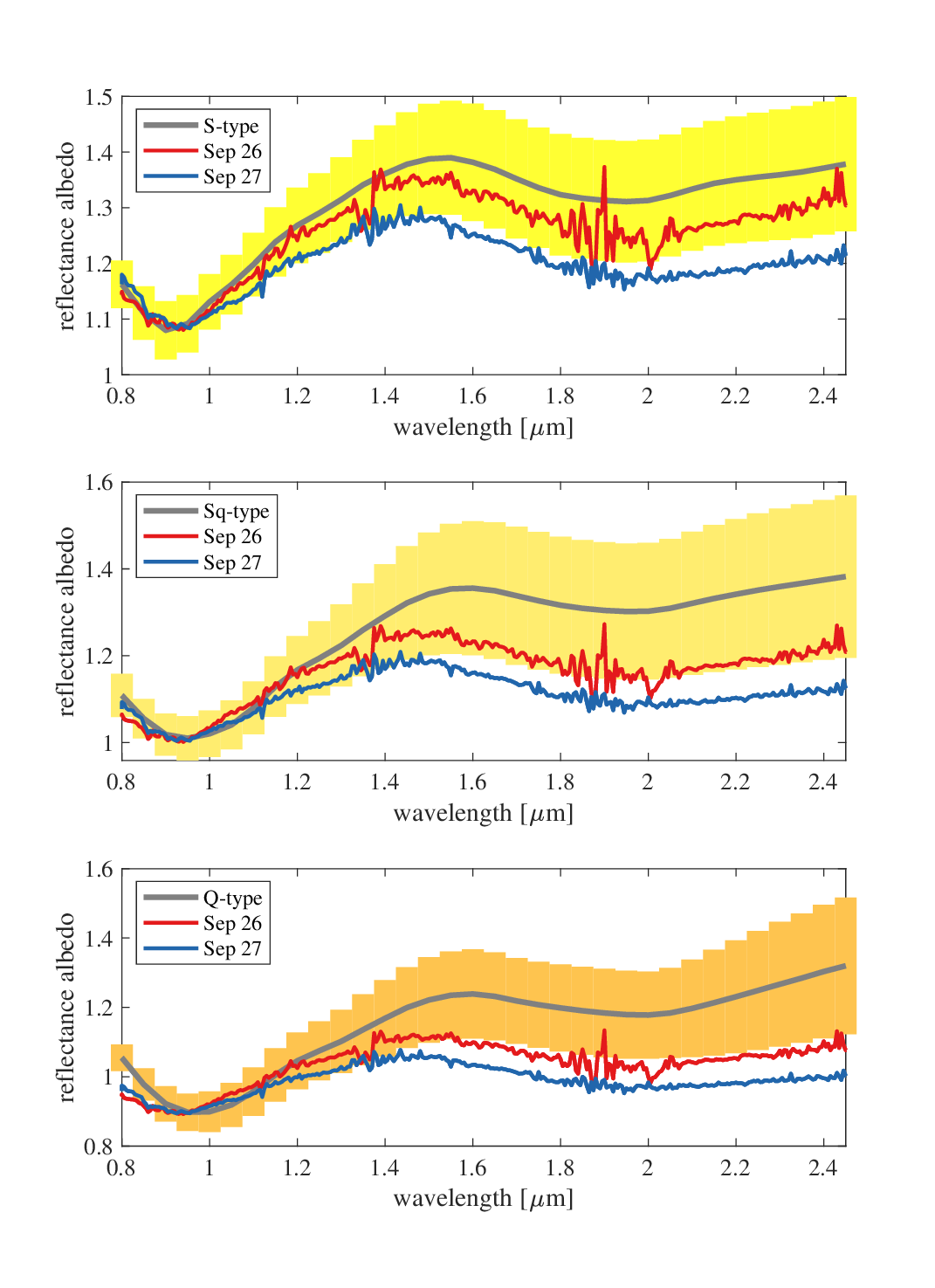}
	\caption{
	Pre- (Sep 26, red line) and post- (Sep 27, blue line) spectra of the Didymos system compared to the $1\sigma$ range (yellowish-orange bars) around averaged taxonomies (grey lines) of S-type (top), Sq-type (middle), Q-type (bottom). The minimum value of a six-order fit to the $1~\mu m$ absorption band of the Didymos system spectra, was normalized to the respective value of each of the taxonomies, in order to ease the spectra comparison by using the $1~\mu m$ absorption band as an anchor. Both pre- and post-impact spectra match well to the S-type spectrum at the $1~\mu m$ absorption band, and do not fit to the Q-type reflectance spectrum, rejecting an hypothesis of an ejecta with fresh-looking material. The overall spectral slope of the pre-impact spectrum fits well to the S-type spectrum; the same value of the post-impact spectrum better fits S-type than Sq- and Q-types.}
	\label{FIG:SpecTax}
\end{figure}

Each of the reflectance spectra (one per night) fit the average S-type classification \citep{DeMeo2009}, as can be seen on Fig.~\ref{FIG:SpecTax}. Other S-complex sub-flavors (Sq-, Q-, Sr-, Sv-, Sa-types) do not match the reflectance spectra due to large differences of the $1~\mu m$ absorption band features, such as depth and width.

\subsection{Principal Component Analysis}
\label{sec:pca}

We used a Principal Component Analysis (PCA) to compare the seven reflectance spectra to each other and to defined taxonomies. For this test we added a visible reflectance spectrum of Didymos available in the literature \citep{deLeon2010}. We calculated the ratio between the visible to each of the near IR spectra from the slopes of the spectra at the range between 0.8 to 0.9 $\mu m$, and normalized the near IR spectra accordingly (by $\sim1.15$) to keep the reflectance spectrum in unity at 0.5 $\mu m$. We also added a few points on the blue side (0.45 to 0.49 $\mu m$), based on the linear slope of the visible spectrum at 0.5 to 0.7 $\mu m$, in order to be able to use the code provided on the SMASS website (http://smass.mit.edu/) that requires a signal from 0.45 to 2.45 $\mu m$.

\begin{figure}
	\centering
             \includegraphics[width=8cm]{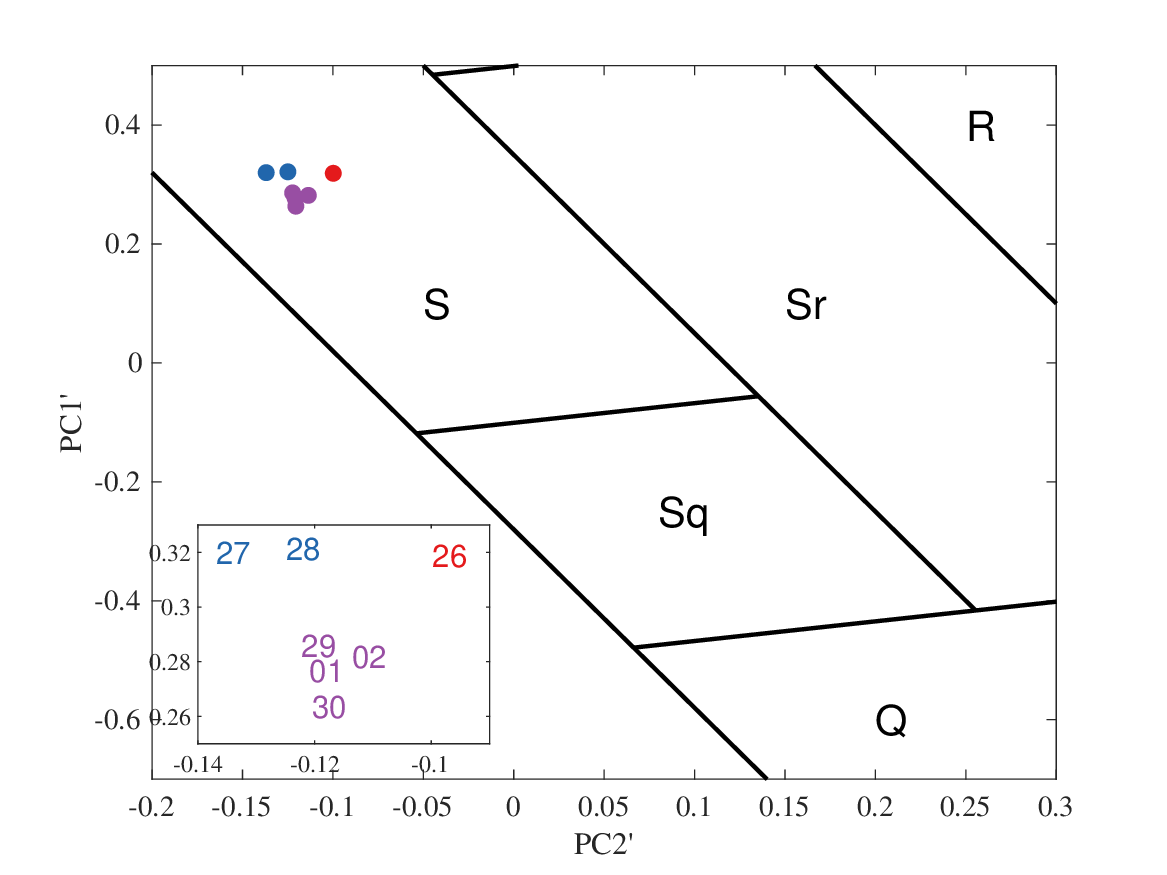}
	\caption{
	PC2'-PC1' plane with different taxonomies limits \citep{DeMeo2009}, and the values of the seven spectra (circles: red Sep 26, blue Sep 27, 28, purple Sep 29, 30, Oct 1, 2). Colors chose as an indication for spectral slope - high (red), mid (purple), low (blue). An inset panel presents a zoom-in plot with the dates of each spectra instead of the dots.
	}

	\label{FIG:PCAmap}
\end{figure}

The PCA analysis shows a clear-cut resemblance to the S-type taxonomy for all seven reflectance spectra (Fig.~\ref{FIG:PCAmap}, with comparison to taxonomy limits on the PCA plan defined by \citealp{DeMeo2009}). Moreover, none of the spectra can be interpret by the PCA analysis as a Q-type classification. There are slight differences between the PCA values of the pre-impact spectrum (measured on Sep. 26), the two spectra that present a lower spectral slope (Sep. 27 and 28) and the last four spectra (Sep. 29, 30, Oct. 1, and 2). It should be noted that the PCA values of the Sep. 27 and 28 spectra are not shifted to the Sq-Q areas on the PC2-PC1 plane, rather to the opposite side. PC1 and PC2 values are presented at Table~\ref{tab:Analysis}.

\subsection{Band Analysis and Mineralogy}
\label{sec:bandanalysis}

Analyzing the $1~\mu m$ and $2~\mu m$ absorption bands also gives S-type values. Here we follow band analysis protocols from the literature. The protocol presented in \cite{DeMeo2014Q} is useful when only the near-IR spectrum is available. It measures band I’s center, depth and width (at $0.85~\mu m$ and at the reflectance peak longward of the band, at $\sim1.5~\mu m$), and other parameters with the original spectral slope (see DeMeo's Fig. 2). There is small, yet continuous change of the bands’ parameters between the measured spectra, where the pre-impact spectrum has a classical S-type values, that changes to an intermediate values of S- to Sq-types for the first post-impact reflectance spectrum, then returns with time to the S-type values. \cite{Polishook2014} uses a band analysis protocol that removes the slope using \cite{Brunetto2006} power-low equation, that was originally used to describe space weathering levels:
\begin{eqnarray}
    W(\lambda)=F(\lambda)exp(C_3/\lambda)
    \label{eq:DeWeathering}
\end{eqnarray}
where F is the fresh reflectance per wavelength $\lambda$, W is the weathered reflectance and the power-law $C_s$ is the extent of the space weathering. This is equivalent to remove the continuum contribution from the spectra as done in other protocols (\citealp[e.g.][]{deLeon2010}).

In order to further derive a mineralogical classification, we also measured the center of the $2~\mu m$ absorption band after the slope removal and applying a polynomial fit of the {\it{n}}-th order to the band. In addition, we measured the Band Area Ratio (BAR), following the classic protocol (\citealp[e.g.][]{deLeon2010, Sanchez2020}). Like in the PCA analysis (see previous section), we added \cite{deLeon2010}'s visible spectrum to have a full Vis-NIR range. Since the highest value derived from SpeX was set to $2.45~\mu m$, we compared the BAR values to values given by \cite{Sanchez2020} that measured BAR values for different wavelength limits. The relevant limit for our comparison is of 0.7 to 2.45 $\mu m$.

After the slope removal (see Fig. 4 in \citealt{Polishook2014}), the bands' parameters mainly remain fixed within the uncertainty range (bottom panel of Fig.~\ref{FIG:Spec_Sep26Oct2}), suggesting any variation of the band analysis are only due to the spectral slope variation.

The values obtained for the band I center ($0.942\pm0.003$) and the band II center ($2.02\pm0.01$) suggest a L or LL-type ordinary chondrite comparing to known meteorites (see Fig. \ref{FIG:HLLL}, \cite{Sanchez2020}, Fig. 4 of \cite{deLeon2010} and as indicated by \citealt{Dunn2013}). The BAR value was less consistent with values ranging from 0.78 to 1.47. This might be partly due to use of an old visible spectrum that does not represent Didymos at the specific times after the DART impact. The lower bottom of this range, in addition to the band I center, better fits the H and L ordinary chondrite values as seen on Fig.~\ref{FIG:HLLL}, and Fig. 3, mid-top panel, at \cite{Sanchez2020}. Moreover, using spectral measurements from the James Webb Space Telescope, \cite{Rivkin2023JWST} determine Didymos is an LL ordinary chondrite type. Therefore, while our analysis of the Didymos' spectral parameters are most consistent with an L-type ordinary chondrite, we note that this result is inconclusive and should be tested by the Hera space mission.

\begin{figure}
	\centering
             \includegraphics[width=8cm]{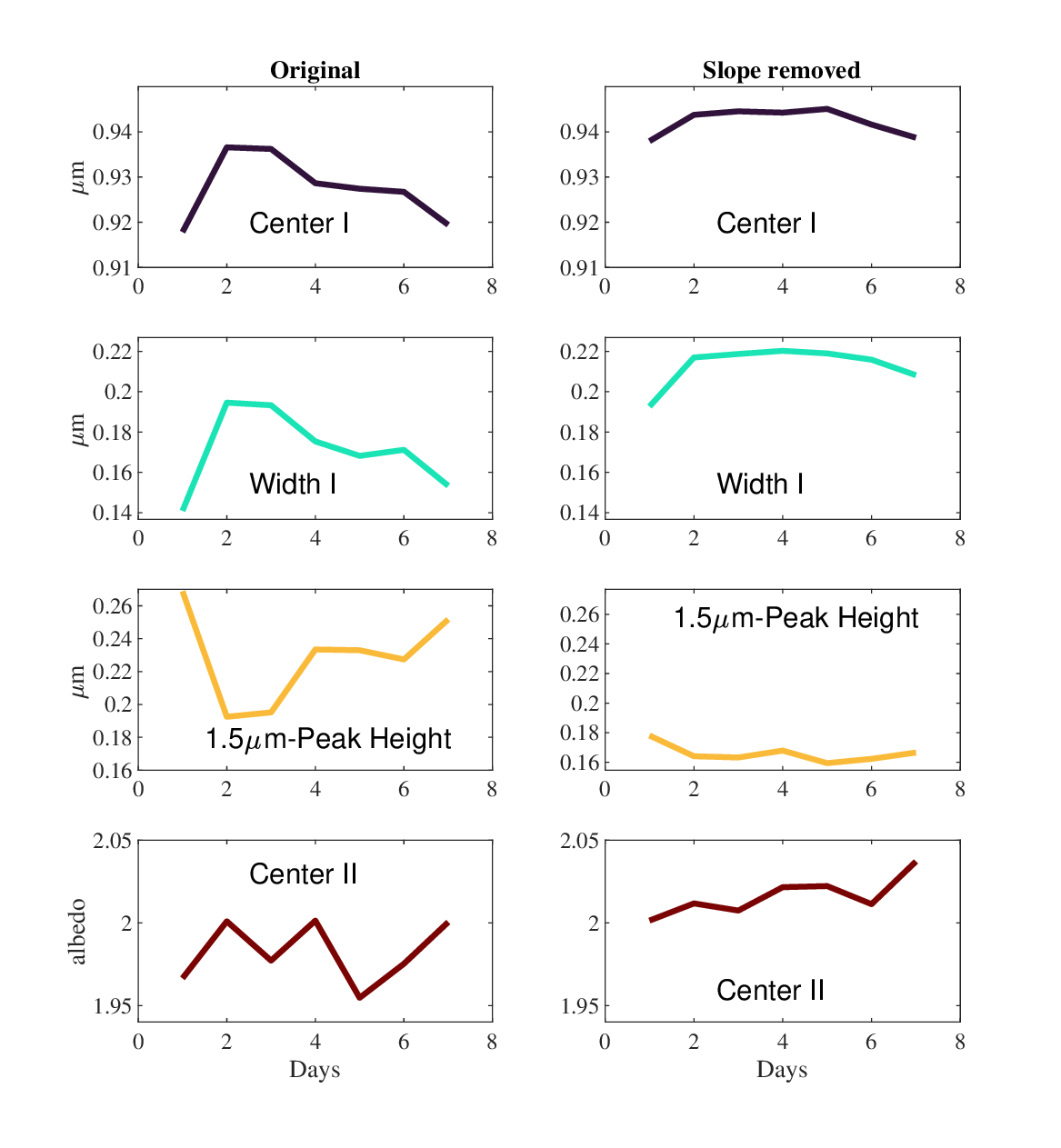}
	\caption{
	Band Analysis. Changes of four parameters with time (Day 1 is Sep 26, day 7 is Oct 2, 2022). Parameters are $1~\mu m$ absorption band's center and width (from $0.85~\mu m$ as described by \citealt{DeMeo2014Q}), the $1.5~\mu m$ height relative to the band I minima (also referred to as 'potential depth' in \citealt{DeMeo2014Q}), and the $2~\mu m$ absorption band's center. The left panels present the band analysis of the original spectra, where relatively larger changes can be measured at the immediate nights after the impact (Sep 27, 28) compared to the pre-impact spectrum and the later spectra. The right panels present the same parameters after the removing of the spectral slope (using \citealp{Brunetto2006} power-low "de-weathering" equation, Eq. \ref{eq:DeWeathering}). The changes in the band parameters almost disappear, suggesting the only band parameter that changed during these nights is the spectral slope and not the other parameters that are affected by composition.
	}

	\label{FIG:BandAnalysis}
\end{figure}

\begin{figure}
	\centering
             \includegraphics[width=8cm]{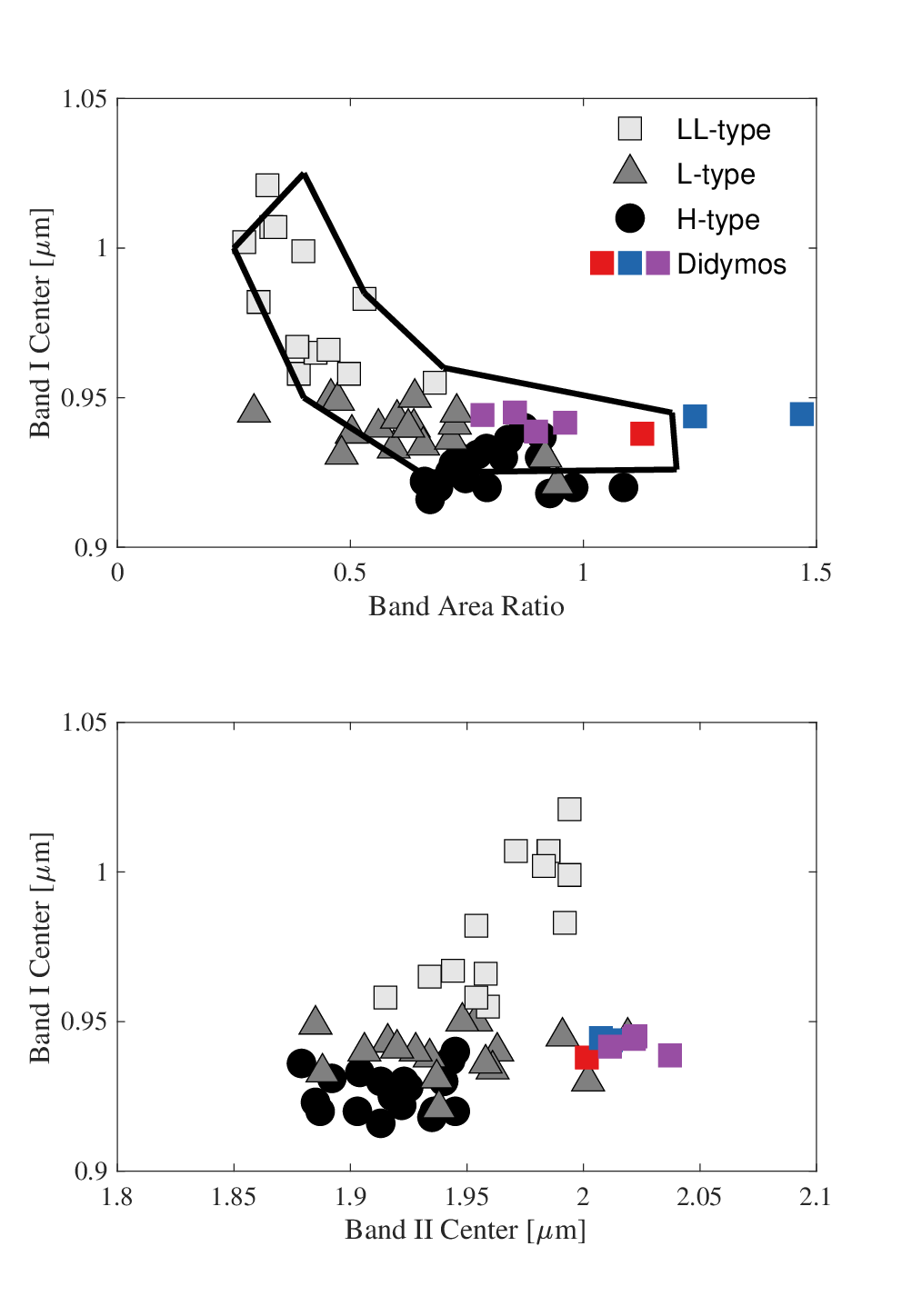}
	\caption{
	Band Analysis. Comparing band I center, band II center and band area ratio of meteorites of H-, L-, and LL-type ordinary chondrite to the Didymos system's values measured by the IRTF. While the band I center and the BAR better fit the H- and L-types, band II center better fits the L- and LL-types. Therefore, while L-type is preferable, the result is inconclusive.
	}

	\label{FIG:HLLL}
\end{figure}

The band analysis values, before and after removing of the spectral slope, are presented at Table~\ref{tab:Analysis} and Fig.~\ref{FIG:BandAnalysis}.

\begin{deluxetable*}{lllllllllllll}
\tablecolumns{13}
\tablewidth{0pt}
\tablecaption{\label{tab:Analysis} 
Spectral analysis}
\tablehead{
\colhead{Night}    &
\colhead{Original}    &
\colhead{Slope 0.8-} &
\colhead{Slope 0.8-} &
\colhead{Cs} &
\colhead{Band I} &
\colhead{Band II} &
\colhead{Band I} &
\colhead{Band I} &
\colhead{Band II} &
\colhead{BAR} &
\colhead{PC1'} &
\colhead{PC2'} \\
\colhead{}       &
\colhead{or de-sloped}       &
\colhead{-2.45 $\mu m^{-1}$}       &
\colhead{-1.5 $\mu m^{-1}$}       &
\colhead{} &
\colhead{center $\mu m$} &
\colhead{center $\mu m$} &
\colhead{width $\mu m$} &
\colhead{area} &
\colhead{area} &
\colhead{} &
\colhead{} &
\colhead{}
}
\startdata
Sep 26 & Original & 0.082 & 0.40 &  & 0.92 & 1.97 & 0.14 & 0.03 & 0.06 & 2.21 & 0.319 & -0.100 \\ 
  & de-sloped &  &  & -0.13 & 0.94 & 2.00 & 0.19 & 0.04 & 0.05 & 1.13 &  &  \\ 
Sep 27 & Original & 0.026 & 0.28 &  & 0.94 & 2.00 & 0.19 & 0.03 & 0.05 & 1.73 & 0.320 & -0.137 \\ 
  & de-sloped &  &  & -0.05 & 0.94 & 2.01 & 0.22 & 0.03 & 0.04 & 1.24 &  &  \\ 
Sep 28 & Original & 0.031 & 0.29 &  & 0.94 & 1.98 & 0.19 & 0.02 & 0.05 & 2.10 & 0.321 & -0.125 \\ 
  & de-sloped &  &  & -0.05 & 0.94 & 2.01 & 0.22 & 0.03 & 0.04 & 1.47 &  &  \\ 
Sep 29 & Original & 0.064 & 0.34 &  & 0.93 & 2.00 & 0.18 & 0.03 & 0.04 & 1.32 & 0.286 & -0.122 \\ 
  & de-sloped &  &  & -0.10 & 0.94 & 2.02 & 0.22 & 0.04 & 0.03 & 0.78 &  &  \\ 
Sep 30 & Original & 0.075 & 0.34 &  & 0.93 & 1.95 & 0.17 & 0.03 & 0.04 & 1.38 & 0.264 & -0.120 \\ 
  & de-sloped &  &  & -0.11 & 0.95 & 2.02 & 0.22 & 0.04 & 0.03 & 0.85 &  &  \\ 
Oct 01 & Original & 0.065 & 0.33 &  & 0.93 & 1.98 & 0.17 & 0.03 & 0.04 & 1.47 & 0.277 & -0.121 \\ 
  & de-sloped &  &  & -0.10 & 0.94 & 2.01 & 0.22 & 0.04 & 0.03 & 0.96 &  &  \\ 
Oct 02 & Original & 0.081 & 0.37 &  & 0.92 & 2.00 & 0.15 & 0.03 & 0.03 & 0.92 & 0.282 & -0.114 \\ 
  & de-sloped &  &  & -0.13 & 0.94 & 2.04 & 0.21 & 0.04 & 0.03 & 0.90 &  & 
\enddata
\tablecomments{Spectral analysis: observing date, original spectrum or after slope removal, linear spectral slope at the ranges of 0.8 to 2.45 and 0.8 to 1.5 $\mu m$, power low $C_s$ of the "de-weathering" equation \ref{eq:DeWeathering}, $1~\mu m$ band's center, $2~\mu m$ band's center, $1~\mu m$ absorption band's width (at $~0.85\mu m$, according \citealt{DeMeo2014Q}), $1~\mu m$ band's area, $2~\mu m$ band's area, band area ratio (BAR), principal components analysis parameters PC1', PC2'.}
\end{deluxetable*}

\section{Discussion \& Conclusions} \label{Sec:DiscussionConclusion}

\subsection{Ejecta Cloud as Representative of Dimorphos' material}
\label{sec:SpecOfDimorphosCloud}

Following the DART impact, the brightness of the Didymos system increased due to the large cross section of the slow ejecta cloud \citep[e.g.][]{Thomas2023, Moreno2023, Li2023, Graykowski2023}. While before the impact Dimorphos’ contribution was only $\sim5\%$ of the shared light of the system (\citealt{Scheirich2022}), after the impact the ejecta became a significant source of light. Assuming a similar albedo, we can estimate the fraction of light reflected from Didymos and the ejecta cloud, during the spectral measurements, from the overall increase in magnitude. The Post-impact brightness of the Didymos system was estimated using the photometry measured by the $1.54~m$ Danish telescope at La-Silla and was submitted for publication by (Rozek et al. private communication). We used the published photometric measurements with the smallest aperture ($1.5''$), to have a better fit to the $0.8''$ width of SpeX's slit. The brightness of the system at the times of the spectral measurements, was estimated by a linear fit to the relevant area of the lightcurve. Using a simple correlation between the change in measured light ($\Delta Mag$) and the combined surface area of Didymos, Dimorphos and the ejecta cloud ($S_{index}$)
\begin{eqnarray}
    \Delta Mag=-2.512log\frac{S_{Didymos}+S_{Dimorphos}+S_{ejecta}}{S_{Didymos}+S_{Dimorphos}},
    \label{eq:EjectaRatio}
\end{eqnarray}

the fraction of the ejecta cloud's contribution to the overall light can be estimated (Table~\ref{tab:Brightenning}). We find that during our first post-impact spectral measurement (Sep 27) the light from the ejecta cloud consist a large fraction of about $\sim64\%$, and this value decreases to $\sim23\%$ at our last measurement (Oct 2). Therefore, the derived high fraction during the first few nights, allows us to treat the measured spectra as a true representative of the material ejected from Dimorphos, and not just of the surface of Didymos.

This understanding, and the large similarity between the pre- and post-impact spectra, strongly support the notion that Dimorphos is equivalent to Didymos in composition and spectral behaviour. This is a unique and novel result that was never obtained before. It strongly supports binary asteroid formation models that include breaking up of a single body, due to rotational breakup of km-wide bodies \citep[e.g.][]{Scheeres2007, Walsh2008}). It also coexists with dynamical family asteroids that have similar taxonomies \citep[e.g.][]{Masiero2015}. It also suggests that the surface composition of some asteroids, most probably those that are similar in nature (i.e. composition, size, structure) to Dimorphos, represents the composition of their inner layers. This result should be added to a long list of the DART mission successes.

\subsection{Spectral Slope Variation Due to Grain Size Variation}
\label{sec:DimorphosCloudGrainSize}

The spectral slope variation, mainly seen on Sep 27 and Sep 28 ($T_0+14$, $T_0+38$ hours after impact), might be explained by three physical processes: compositional variation of the ejected material; ejection of "fresh", non-weathered material; or ejected material with varied grain size.

Composition variation from the ejected material should be refuted on the basis of the band analysis that is resulted with almost identical parameters for all seven reflectance spectra (Table~\ref{tab:Analysis}, Fig.~\ref{FIG:Spec_Sep26Oct2} lower panel and Fig.~\ref{FIG:BandAnalysis}). A different material, even within the larger group of ordinary chondrite, should present different band parameters (such as the center of the $1 \mu m$ and $2 \mu m$ absorption band; \citealp{Dunn2010}).

Excavated non-weathered material by the DART impact was expected ahead of the impact, since weathered layers on the Moon and asteroid Itokawa were measured to be only nano- to micro-meter thick \citep[e.g.,][]{Chapman2004, Abe2006}, and the impact was expected to remove, and indeed it removed, much thicker layer (As observed by the LICIACube CubeSat, (\citealt{Dotto2023}); by direct images of boulders in Dimorphos' vicinity (\citealt{Jewitt2023}); and by measuring the formed dust tail (\citealt{Moreno2023})). Leading theories suggest that the surfaces of Q-type asteroids became fresh due to a "gardening" process following planetary encounters with planets, a fast spin, minor impacts, etc (\citealt{DeMeo2023, Binzel2010, Nesvorny2010} and references within). Relevant supporting observations are the cases of active asteroids P/2010 A2, (6478) Gault, and some asteroid pairs (see section \ref{sec:intro} for details).

However, the shallow-slope spectra of the Didymos system do not present Q-type spectral features, especially not the longer-wavelength band center, depth and width (Fig. \ref{FIG:SpecTax}). The spectral slope, including those with a smaller value (from Sep 27 and Sep 28), still fit better to the slope of the S-type classification. Therefore, it should be concluded that the subsurface material excavated by the DART impact is not "fresh" as defined by fresh lunar material or by Q-type asteroids. 

This might be the result of Dimorphos formation from material that was resting on top of Didymos for many years, became weathered over time, and later on was ejected by Didymos' fast spin and re-accumulated in orbit into Dimorphos. In this scenario the DART impact released already-weathered particles that do not present Q-type spectrum. If true, asteroid satellites of the ordinary chondrite class can not be characterised as Q-types since they formed from already weathered material. Such a behaviour should be tested in future research.

We should add a caveat, stating that while the scenario described above matches the behavior of LL ordinary chondrite type, it less fit to the H- and L-types cases, since the spectra of these types are less altered by space weathering possibly due to a different rate of weathering between olivine-rich and pyroxene-rich asteroids \citep{Brunetto2015}. In this case the band width and depth will be narrow for both weathered and non-weathered asteroid surface, and the difference in spectral slope seems to span a wide range due to multiple factors as described in section \ref{sec:intro} and in \cite{Brunetto2015}.

Refuting slope variation by composition and weathering levels, makes the grain size alternative seems more plausible. Models already suggested that particle size primarily affects the slope of an absorption feature and overall reflectance \citep{2001JGR...10610039H}. Laboratory experiments showed that larger particle size typically means lower spectral slope (\citealp{Hasegawa2019}, \citealp{Reddy2015} and references within). Moreover, the $1~\mu m$ absorption band's center position is insensitive to the grain size \citep{Cloutis2015}, supporting the notion of grain size-dependent spectral variation, since it explains why the slope changes while the band center remains fixed.

The level of spectral variation we measured is also seen on meteorite samples. In order to better understand how differences in grain size can influence the spectral slope, we analyzed the near-infrared spectra ($0.8-2.45~\mu m$) of the ordinary chondrites Chelyabinsk obtained by \cite{Bowen2023}. Chelyabinsk's reflectance spectrum was identified as a closer match to Didymos by \cite{Rivkin2023JWST}. The spectrum was obtained for five different grain size groups (45-90, 90-150, 150-300, 300-500, and 500-1000 $\mu m$) using an Analytical Spectral Devices LabSpec 4 HR spectrophotometer. 
The spectrum was normalized at the minimum value of a 6-order fit to the $1~\mu m$ absorption band. The spectral slope was calculated from a linear fit to the data from $0.8~\mu m$ to the reflectance maximum at $\sim1.5~\mu m$. The reflectance maximum at $0.8~\mu m$ was determined by fitting a straight line from 0.8 to 0.84 $\mu m$, and the maximum at $\sim1.5~\mu m$ was determined by fitting a third-order polynomial around this wavelength. A total of 50 measurements of the spectral slope were obtained by sampling slightly different data points. The final value is given by the average of all measurements and the uncertainties correspond to the standard deviation of the mean.

\begin{figure}
	\centering
             \includegraphics[width=8cm]{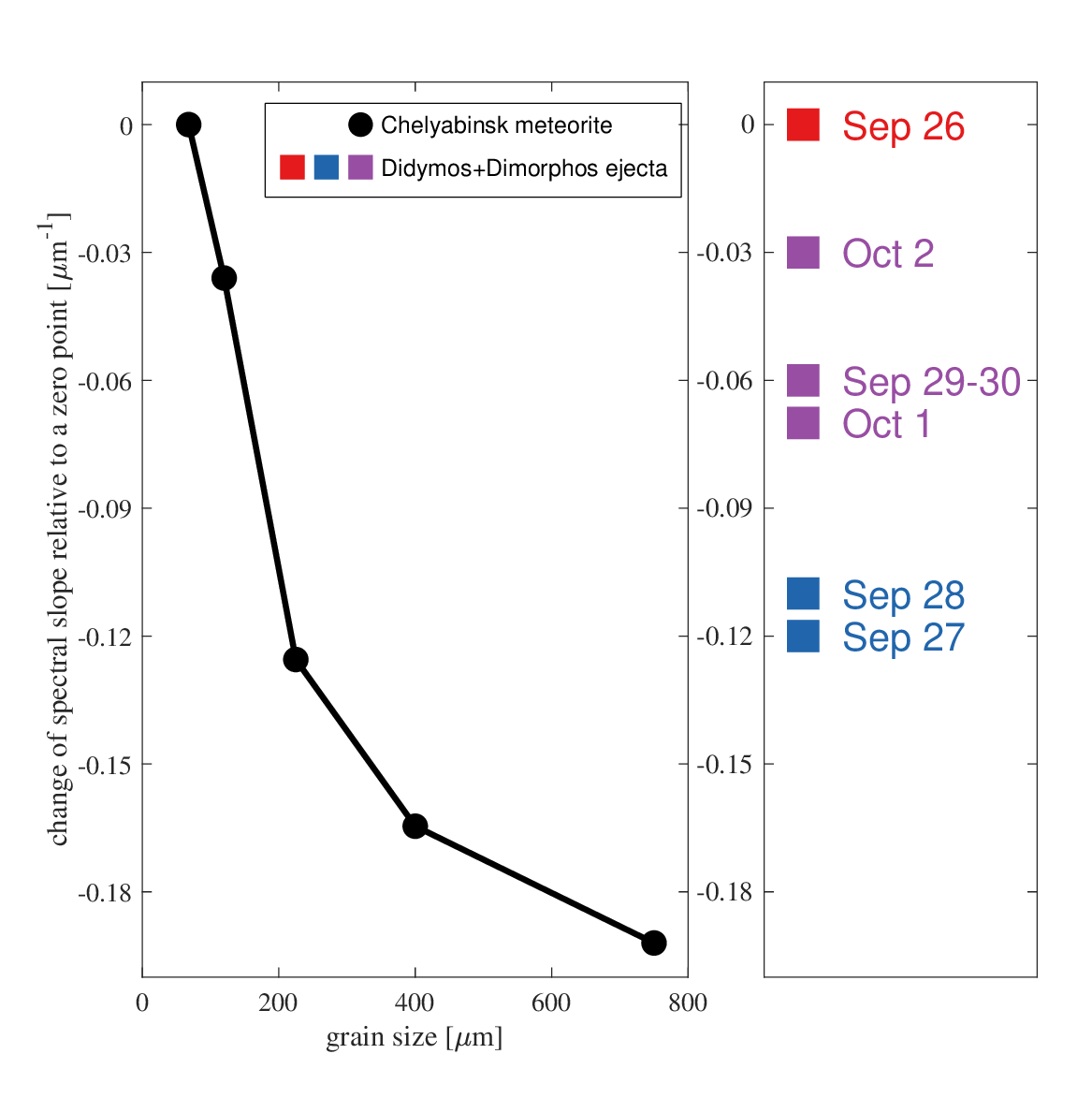}
	\caption{
	Differences between spectral slopes of the Chelyabinsk meteorite's grains of different sizes, relative to the spectral slope of grains in the size of $45-90~\mu m$ (left, black circles), compared to differences between spectral slopes of the Didymos system after the DART impact, relative to the spectral slope at pre-impact (right, colorful squares, same color scheme as in Fig.\ref{FIG:PCAmap}). Both Y-axes are to scale showing the similar range of spectral variation. Uncertainty is smaller than the markers size.
	}
	\label{FIG:GrainSize}
\end{figure}

The measurements show that decreasing spectral slope is correlated with increasing grain size. For a valid comparison, we used the same wavelength range of $0.8-1.5~\mu m$ for calculating the spectral slope of Didymos (Values presented in Table \ref{tab:Analysis}). We calculated a change in slope from fine grains ($45-90~\mu m$) to large grains ($500-1000~\mu m$) to be $\sim 0.2~\mu m^{-1}$. This range is equivalent to the change in slope between the pre-impact spectrum to the immediate post-impact spectrum (Sep 27 and Sep 28) which is $\sim 0.12~\mu m^{-1}$. The changes in spectral slope of the other spectra (Sep 29 to Oct 2) are smaller, and fit smaller slope changes seen for smaller grains as seen at Fig. \ref{FIG:GrainSize}.

\begin{figure}
	\centering
             \includegraphics[width=8cm]{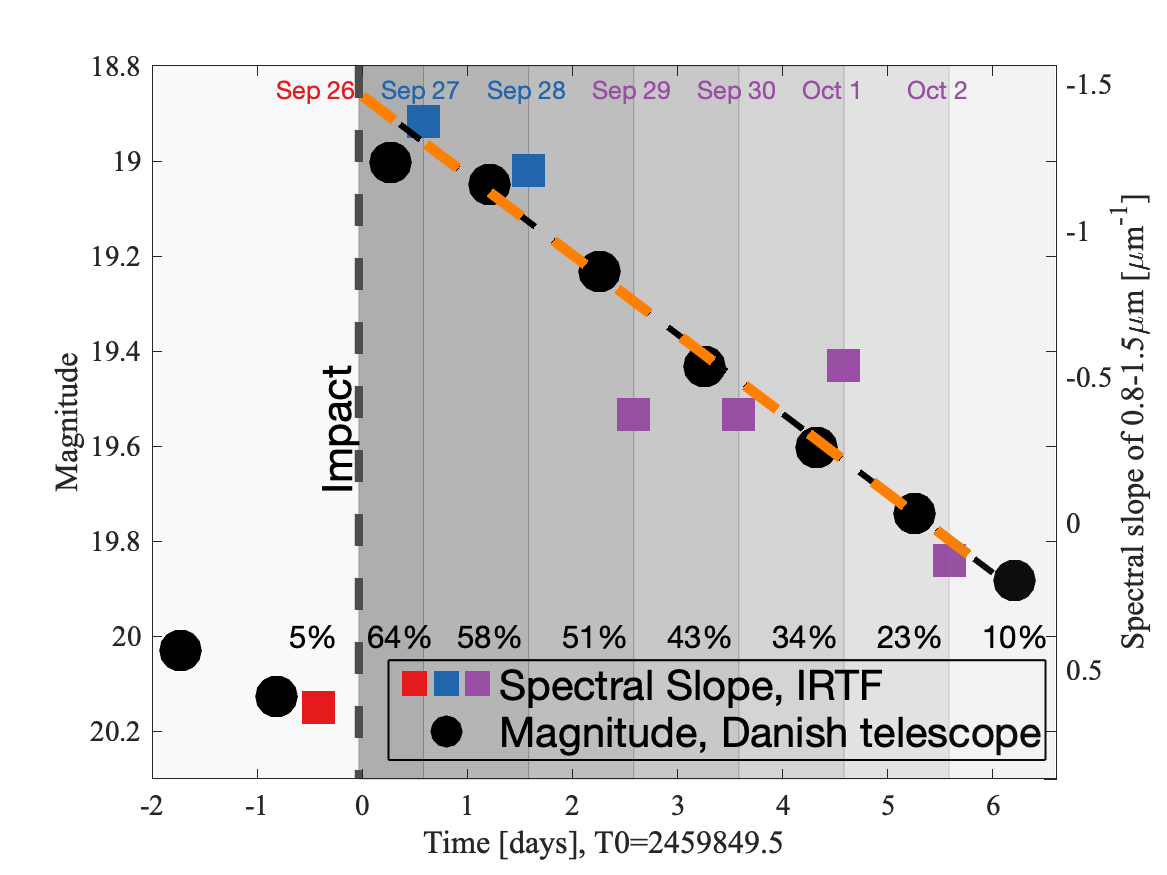}
	\caption{
	Spectral slope measured by IRTF (colorful squares, same color scheme as in Fig.\ref{FIG:PCAmap} and Fig.\ref{FIG:GrainSize}) compared to the changing brightness of the Didymos system (in absolute magnitude values in the V-band, usually marked as $H_v$) following the impact as measured by the Danish 1.54 m telescope (black circles; Rozek et al. submitted for publication) The linear slopes of the two parameter sets were drawn at the same scale using a coefficient of 10.27 and a constant of 16.04 (black and orange dashed-lines). The photometry of the Danish telescope was used for its 1.5'' aperture, which is the most equivalent to the IRTF/SpeX 0.8'' slit-width. The photometric error ($\sim0.04~mag$) is smaller than the size of the symbols. The transparent background represents the decreasing contribution of the ejecta from Dimorphos to the overall light. The used transparency value (written as percentage at the bottom of each section) is the ratio calculated from the Danish telescope photometry. The uncertainty of these ratios are $\sim5\%$. Before the impact, Dimorphos surface is about $5\%$ of Didymos surface.
	}
	\label{FIG:Phot_Spec_Time}
\end{figure}

In order to explain the grain-size variation that might caused the spectral slope variation, we need to understand the size distribution of the ejecta grains, and follow the system evolution after the impact. \cite{Moreno2023} estimated that the particles size distribution ranges from $1~\mu m$ to $5~cm$, a scope determined by a power law index of -2.7 (up to $3 ~mm$, larger particles are control by a steeper index of -3.7) that best fits the photometric observations. Models suggested that solar radiation pressure pushes away ejecta particles smaller than $100~\mu m$ in less than 10 hours \citep{Li2023}, before our first post-impact observation (at $\sim{T_0+14}$). Indeed, images taken by the Hubble Space Telescope show the formation of hundreds of km-long dust tail ahead of our first post-impact observation (Fig. 4, panel d at \citealt{Li2023}). On the first two observing runs ($T_0+14$ and $T_0+38$), the ejecta cloud, with its $\geq100~\mu m$ particles, contributes significant amount of light compares to Didymos (at a level of $60-70\%$), thus the spectral slope is lower as observed. Indeed, the first mutual event detection was detected with photometry at $\sim T_0+29~h$, suggesting first clearing of the ejecta at this time and forward \citep{Thomas2023, Moskovitz2022}. However, smaller particles are still available on the nearby Didymos surface, and their contribution to the overall flux increases with time as the ejecta cloud dispersed, and the system brightness reduces (Table~\ref{tab:Brightenning}). Drawing the decreasing brightness of the Didymos system and the increasing spectral slope by time, presents a similar behavior, with a coefficient of 10.27 between the two linear slopes (Fig.\ref{FIG:Phot_Spec_Time}). This suggests that as the ejecta cloud dissipates with time, the light reflected from the Didymos surface dominates larger fraction of the entire light, thus the peak of the particle size distribution is shifted back to smaller sizes.
This supports the notion that the spectral slope of the Didymos system is indeed controlled by the grain size. Moreover, it shows that Didymos surface is dominated by grains smaller than $100~\mu m$. The Hera space mission will be able to confirm this prediction. The temporal change of the reflectance spectrum of the Didymos system also suggests that if dust particles arrived from Dimorphos to Didymos, they were not significant in order to change its spectrum for good.

\begin{deluxetable*}{llcc}
\tablecolumns{4}
\tablewidth{0pt}
\tablecaption{\label{tab:Brightenning} 
Brightness change of the Didymos system}
\tablehead{
\colhead{Night}    &
\colhead{$T_{spec}$}    &
\colhead{$\Delta Mag$} &
\colhead{\Large{$\frac{S_{ejecta}}{S_{Didymos}+S_{Dimorphos}+S_{ejecta}}$}} \\
\colhead{}       &
\colhead{}       &
\colhead{}       &
\colhead{}
}
\startdata
Sep 26, 2022 & 2459849.0833 & 0 & $0\%$ \\ 
Sep 27, 2022 & 2459850.0833 & -1.12 & $64\%$ \\ 
Sep 28, 2022 & 2459851.0833 & -0.95 & $58\%$ \\ 
Sep 29, 2022 & 2459852.0833 & -0.78 & $51\%$ \\ 
Sep 30, 2022 & 2459853.0833 & -0.62 & $43\%$ \\ 
Oct 01, 2022 & 2459854.0833 & -0.45 & $34\%$ \\ 
Oct 02, 2022 & 2459855.0833 & -0.28 & $23\%$ \\ 
\enddata
\tablecomments{Brightness change of the Didymos system by time of the spectral measurements compared to a pre-impact baseline (20.1 mag), and ejecta cloud ratio, based on photometric measurements collected by the Danish 1.54 m telescope at La Silla, using aperture of 1.5'' (Rozek, private comm.). The uncertainty of the photometry is $\sim0.04~mag.$}
\end{deluxetable*}

\section{Summary} \label{Sec:Summary}

We conducted carefully calibrated near-IR spectral observations of the Didymos system, on the day before, and the six days following the DART impact. We find that:

- Overall, the spectral signature of the Didymos system remained fixed (S-type classification, with some indication for an L-type ordinary chondrite) during the week of observations.

- The high brightening of Didymos following the impact, indicates the Dimorphos ejecta dominated this brightening at a level of $60-70\%$. This confirms both Didymos and Dimorphos are constructed from the same silicate material, supporting binary formation models of a single source.

- The spectral slope of the system decreased following the impact, and almost completely returned to its original level after six days.

- No Q-type like spectra was measured, which is not coexist with the notion that non-weathered material is located below the nano- to micro-meter wide top layer of the surface. Perhaps this is the standard case for asteroid satellites of the ordinary chondrite class, that are formed by accumulation of already weathered material, that is ejected due to a fast rotation of the main asteroid.

- The change in spectral slope, while the $1~\mu m$ and $2~\mu m$ absorption band parameters mostly remained fixed, suggesting a change in the grain size dominating the light from the Didymos system, that is first controlled by the larger particles ejected from Dimorphos (after smaller particles are carried away by solar radiation pressure), and later controlled by the smaller particles on Didymos' surface as the ejecta cloud disperses. This spectral slope change coexist with laboratory measurements of meteorites at a range of grain size.

- The Didymos surface are most probably covered with fine grains, smaller than $100~\mu m$, that dominates the spectral behaviour. This prediction can be tested by the Hera space mission to the Didymos system.

\section{Acknowledgements}
  We thank Richard Binzel and Jessica Sunshine for helpful discussions that are always insightful. We appreciate Agata Rozek for providing us her Danish telescope photometry of the Didymos system around the DART impact. We thank the IRTF staff, especially telescope operators Dave Griep and Greg Osterman for their careful, yet fast, "driving" of the IRTF's 3m telescope during the tense hours of saving the world. We are grateful to Ping Chen and Yuri Beletsky for their aid with spectral data collected by the Magellan/FIRE spectrograph, that did not make it to the paper. DP is thankful to the Israel Space Agency and their support through the IAWN collaboration. DP is also thankful to the DART mission staff and investigation team that made it possible for him to have an intense, insightful and fun astronomical week of his life at the end of September 2022.
  
  Observations reported here were obtained at the NASA Infrared Telescope Facility, which is operated by the University of Hawaii under contract 80HQTR19D0030 with the National Aeronautics and Space Administration. The MIT component of this work is supported by NASA grant 80NSSC18K1004 and 80NSSC18K0849. The work of JS and VR was supported by NASA Near-Earth Object Observations grant NNX17AJ19G.

%

\vspace{5mm}

\bibliography{main.bib}
\bibliographystyle{aasjournal}

\end{document}